\documentclass[12pt,preprint]{emulateapj}

\def\be{\begin{equation}}
\def\ee{\end{equation}}
\def\bea{\begin{eqnarray}}
\def\eea{\end{eqnarray}}

\usepackage{graphicx}

\begin{document}

\title{The lightcurve of the macronova associated with the long-short burst GRB 060614}
\author{Zhi-Ping Jin$^{1}$, Xiang Li$^{1,2}$, Zach Cano$^{3}$, Stefano Covino$^{4}$, Yi-Zhong Fan$^{1}$, and Da-Ming Wei$^{1}$}
\affil{
$^1$ {Key Laboratory of dark Matter and Space Astronomy, Purple Mountain Observatory, Chinese Academy of Science, Nanjing, 210008, China.}\\
$^2$ {University of Chinese Academy of Sciences, Yuquan Road 19, Beijing, 100049, China.}\\
$^3$ {Centre for Astrophysics and Cosmology, Science Institute, University of Iceland, 107 Reykjavik, Iceland.}\\
$^4$ {INAF/Brera Astronomical Observatory, via Bianchi 46, I-23807 Merate (LC), Italy.}
}
\email{yzfan@pmo.ac.cn (YZF)}

\begin{abstract}
The {\it Swift}-detected GRB 060614 was a unique burst that straddles an imaginary divide between long- and
short-duration gamma-ray bursts (GRBs), and its physical origin has been heavily debated over the years.
Recently, a distinct very-soft F814W-band excess at $t\sim 13.6$ days after the burst was identified
in a joint-analysis of VLT and HST optical afterglow data of GRB~060614, which has been interpreted as evidence for an accompanying Li-Paczynski macronova (also called a kilonova).
Under the assumption that the afterglow data in the time interval of $1.7-3.0$ days after the burst are due to external forward shock emission,  when this assumption is extrapolated to later times it is found that there is an excess of flux in several multi-band photometric observations.
This component emerges at $\sim$4 days after the burst, and it may represent the first time that a multi-epoch/band lightcurve of a macronova has been obtained.
The macronova associated with GRB 060614 peaked at  $t\lesssim 4$ days after the burst, which is significantly earlier than that observed for a supernova associated with a long-duration GRB.
Due to the limited data, no strong evidence for a temperature evolution is found. We derive a conservative estimate of the macronova rate of $\sim 16.3^{+16.3}_{-8.2}~{\rm Gpc^{-3}}{\rm yr^{-1}}$, implying a promising prospect for detecting the gravitational wave radiation from compact object mergers by upcoming Advanced LIGO/VIRGO/KAGRA detectors (i.e., the rate is ${\cal R}_{\rm GW} \sim 0.5^{+0.5}_{-0.25}(D/200~{\rm Mpc})^{3}~{\rm yr^{-1}}$).
\end{abstract}

\section{Introduction}
It is widely accepted that the merger of a binary compact object system (either a neutron-star, NS, binary, or a stellar-mass black hole, BH, and NS binary) produces the high-energy $\gamma$-ray emission in a short-duration gamma-ray burst (SGRB) event \citep{Eichler1989,Narayan1992,Berger2014}. Indirect evidence for SGRBs originating from compact-binaries \citep[][]{Gehrels2005,Fong2010,Leibler2010,Fong2013,Berger2014} include the location of SGRBs in elliptical galaxies, no associated supernova (SN), large galaxy offsets ($>100$ kpc) that match population synthesis predictions for compact binaries, and weak spatial correlation of SGRBs and regions of star formation within their host galaxies (when the hosts can be unambiguously identified).

A ``smoking-gun" signature for the compact-binary origin of an SGRB would be the detection of the so-called Li-Paczynski macronova (also called a kilonova),
which is a near-infrared/optical transient powered by the radioactive decay of $r$-process material synthesized in the ejecta that is launched during the merger event \citep[e.g.,][]{Li1998,Kulkarni2005,Rosswog2005,Metzger2010,Korobkin12,Barnes2013,Kasen2013,Tanaka2013,Tanaka2014,Grossman2014,Kisaka2015a,Kisaka2015b,Lippuner2015}.
Macronovae are expected to peak in infrared bands, and they display very soft spectra.
As such, macronova signals are very hard to detect. A breakthrough was made in June 2013. In the late afterglow of the canonical SGRB 130603B ($z=0.356$), an infrared transient was interpreted as a macronova produced during a compact-binary merger \citep{Tanvir2013,Berger2013,Hotokezaka2013,Piran2014}. Very recently, a significant F814W-band excess component was reported in a re-analysis of the late time optical afterglow data (Yang et al. 2015; Y15 hereafter) of the peculiar event GRB 060614 that shares some properties of both long-duration and short-duration GRBs \citep{Gehrels2006,Fynbo2006,DellaValle2006,Gal-Yam2006,Zhang2007}. The photometric spectral energy distribution (SED) of the excess component is so soft (the effective temperature is below 3000 K) that a very-weak SN origin has been strongly disfavored. Instead, the excess flux can be interpreted as a macronova powered by the merger of a stellar-mass BH with an NS (Y15).

To date, the published literature regarding photometric evidence of macronovae associated with SGRB 130603B and long-short burst GRB 060614 are based on only a single datapoint in each event\footnote{In Fig. 1 of Y15, at $t\sim 7.8$ days after the burst there was an $I$-band data point that was in excess of the extrapolated power-law decline of the hypothesized forward shock emission. However, its significance was below $3\sigma$.}. To better reveal the physical processes giving rise to the macronova emission, multi-band photometric (and ideally spectroscopic) observations of the transient are needed in order to compare observations with theoretical predictions \citep[e.g.,][]{Barnes2013,Tanaka2013,Hotokezaka2013,Tanaka2014,Grossman2014}.
We therefore revisited the extensive data set of GRB~060614 obtained with the Very Large Telescope (VLT) and Hubble Space Telescope (HST) to produce multi-band lightcurves and SEDs, and use these information to provide evidence, and constrain the nature, of the accompanying macronova.

This work is structured as follows: In Sec. 2 we first review the basic assumptions made in Y15.  Next, we discuss the
necessity/feasibility of relaxing these assumptions, and then extract the lightcurve of the associated macronova. The rate of macronova/compact-object mergers is estimated in Sec. 3, and our results and discussion are presented in Sec. 4.

\section{Extracting the lightcurve of macronova associated with GRB 060614}
To robustly establish the existence of a distinct HST F814W-band excess in the late afterglow of GRB 060614,
Y15 assumed that all of the VLT data were due to the forward shock (FS) and subsequently fitted the $VRI$ data at $t>1.7$ day with the same decline rate. In such an approach, only one $F814W$-band point at about 13.6 days was found to be more than 3$\sigma$ in excess of the fitted FS emission. However, the fitted residuals in Fig. 1 of Y15 display an interesting general trend: the earlier data ($t<4$ days) were usually negative (with respect to the FS afterglow model), while the later data were positive, indicating that the intrinsic FS emission decline was likely steeper than that assumed in their model, and there was likely to be an excess of emission at times earlier than 13.6 days. On the other hand, in numerical simulations, macronova {\it optical} emission usually peaks in a few days to a week (rest frame) after the merger event, and its subsequent contribution to the afterglow emission can be non-negligible \cite[e.g.,][]{Barnes2013,Tanaka2013,Tanaka2014}.
After having solidly established the existence of an excess of flux in the analysis performed by Y15,
we sought to improve the analysis by considering a possible time evolution of the macronova component and modelling the entire afterglow dataset accordingly.

GRB afterglows are expected to be powered by FSs that produce synchrotron emission, which have a power-law like behavior in both time and frequency, where the temporal and energy spectral indices, $\alpha$ and $\beta$, respectively, are defined by $f_{\nu} \propto t^{-\alpha}\nu^{-\beta}$, where $t$ is the time since the GRB was first detected by a satellite \citep[e.g.,][]{Piran2004,Kumar2015}.
For SGRBs,
the afterglow emission emitted after several hours should consist of radiation coming from both the FS and the associated macronova. Hence a macronova lightcurve can, in principle, be ``self-consistently" obtained through a joint fit of the observational data. A key outstanding problem is that current theoretical macronova calculations still suffer from significant uncertainties.
For example, the role of radioactive heating due to the fission of  heavy $r$-process nuclei to the energy deposition rate at, for example, $t\sim 10$ days after the merger, is still poorly understood \citep[e.g.,][Hotokezaka 2015 private communication]{Korobkin12,Wanajo2014}. Moreover, the poorly-constrained electron fraction ($Y_{\rm e}$), the escape velocity distribution and the anisotropy of the outflow play additional roles in shaping the macronova emission \citep{Tanaka2014,Lippuner2015}, are all caveats that should be considered when interpreting our results.

In this work we extracted the possible macronova emission by decomposing the FS emission from the observational data.
A reliable estimate of the FS emission is very crucial, thus the following facts were taken into account: (i) there was a jet break around 1.4 days \citep{DellaValle2006, Mangano2007, Xu2009},
hence data after this time need only be considered; (ii) at $t\sim 1.7-1.9$ days after the burst, the optical to X-ray spectrum is well described by a single power-law \citep{DellaValle2006,Mangano2007,Xu2009}, suggesting that any macronova contribution to the observed flux is negligible; (iii) in the interval of $1.7-3.0$ days after the burst there were two measurements in VLT $VI$ bands and three measurements in VLT $R$ band, allowing us to obtain a relatively reliable estimate of the  FS emission decline. Therefore in this work  we adopt the VLT and HST observational data reduced in Y15,  but we assumed that only the VLT data in the interval of $1.7-3.0$ days are due to only FS emission, and we used these data to determine the single power-law decline of the afterglow.

The observed magnitudes were first corrected to the magnitudes in the $R$ band, assuming an Galactic extinction of $A_{V}$=0.07 mag (Schlegel et al. 1998; Schlafly \& Finkbeiner 2011), the extinction of the host galaxy is SMC like $A_{V}$=0.05 mag, and the intrinsic afterglow spectrum is well described by a single power-law with $\beta=0.81\pm0.08$, as based on the optical and UV data at 150ks fitted by \citet{Mangano2007} and confirmed by \citet{Xu2009}. The fit to the VLT data collected in the time interval of $1.7-3.0$ days yields $\alpha=2.55\pm0.09$.
This is steeper than the decay index of $\alpha=2.30\pm0.03$ obtained in Y15 by assuming all VLT data were FS emission. Such a difference is reasonable/expected since the ``underlying" macronova emission contributes to observations at later epochs, thus causing the LCs to appear to decay at a slower rate. In the slow cooling of a jetted outflow with significant sideways expansion, when the observational frequency is between the so-called typical synchrotron radiation frequency $\nu_{\rm m}$ and the cooling frequency $\nu_{\rm c}$, the decline and spectral indexes are expected to be $\alpha=p$ \citep[after the jet break,][]{Sari1999} and $\beta=(p-1)/2$ \citep{Piran2004,Kumar2015}. Interestingly the observed $\beta=0.81\pm0.08$ and our inferred $\alpha=2.55\pm 0.09$ are in good agreement with the standard afterglow model (i.e. they both predict an electron index of $p \approx 2.6$).

\begin{figure}
\begin{center}
\includegraphics[width=0.5\textwidth]{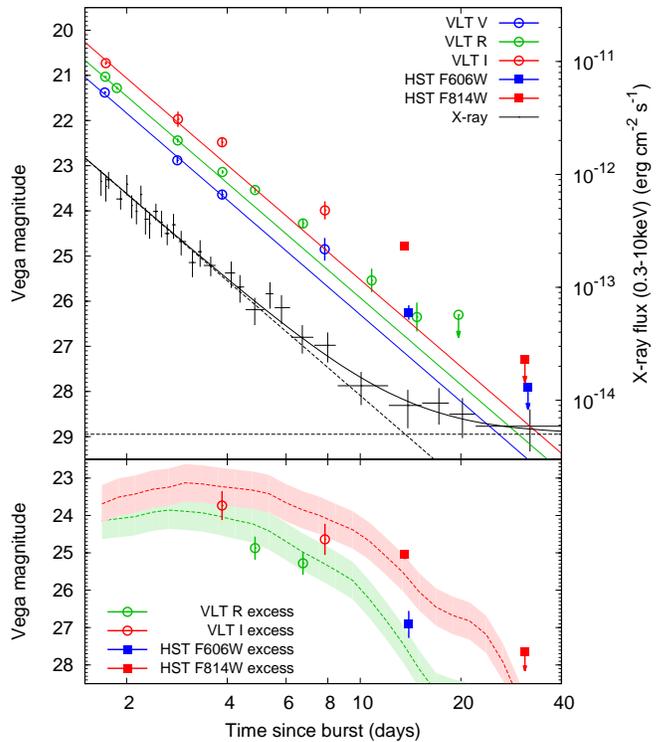}
\end{center}
\caption{The observed lightcurves of the macronova associated with GRB 060614.
\textbf{Top:} The data points are adopted from Y15 but just the VLT data in the time interval of $1.7-3.0$ days
are assumed to arise soley from FS emission, and the solid lines represent the fit  ($\propto t^{-2.55}$).
The simultaneous X-ray emission, retrieved from the UK Swift Science Data Centre \citep{Evans2009}, can be fitted by $t^{-2.55}$ plus a constant flux. A constant X-ray flux of $(8\pm4)\times10^{-15}~{\rm erg}~{\rm cm}^{-2}~{\rm s}^{-1}$ was obtained by \citet{Mangano2007} and was interpreted as the emission from a possible Active Galactic Nucleus, or it was simply a statistical fluctuation because of the low measured flux that was very close to the detection threshold of {\it Swift} XRT. Simultaneous with the very late/weak ``plateau-like" X-ray emission, the HST F814W-band flux drops as $t^{-3.2}$, ruling out a possible energy injection mechanism.
\textbf{Bottom:} Significant excess appears at late times.
Note that the data are not corrected for any extinction, and only ``macronova" emission points with a significance above $2\sigma$ were kept.
The dashed lines, adopted from Y15, are macronova model lightcurves generated from numerical simulation for the ejecta from a BH$-$NS merger,
with a velocity $\sim 0.2c$ and mass $M_{\rm ej}\sim0.1M_{\odot}$, by \citet{Tanaka2014}.
The green and red lines are in $R$ and $I$ bands,  and shadows represent a possible uncertainties of 0.5 magnitudes (Hotokezaka 2015 private communication).
The macronova model is in agreement with the observed data, including those with large uncertainties (i.e. significance below $2\sigma$, see Table 1).
} \label{fig:LC}
\end{figure}

When we subtracted this FS component from the observational data
we found a significant excess in multi-wavelength bands at $t>3$ days,  which may constitute the first multi-epoch/band lightcurve of a macronova ever recorded.
The results are shown in Tab. 1 and Fig. \ref{fig:LC}, where the errors include the uncertainties of the observed magnitudes and the FS model uncertainties.
Although the dataset is still relatively sparse, there is an indication that the macronova emission likely peaked at $t\lesssim 4$ days after the merger event\footnote{After extracting the possible macronova emission from the data,
we find that the macronova was always much fainter than the afterglow between $1.7-3.0$ days after the burst
and its contribution was smaller than the afterglow uncertainties.
Hence our assumption that in the time interval of $1.7-3.0$ days the emission is due to just FS is reasonable.
However, due to the $t^{-2.55}$ decline of the FS emission and the shallow decay of the macronova, at t$\geq$4 days the contribution of macronova to the total flux can not be ignored any longer.}, which is consistent with current numerical simulations \citep[e.g.,][]{Kasen2013,Tanaka2013,Tanaka2014}. In comparison, among GRB-associated SNe, SN 2010bh had the most rapid rise to maximum brightness, peaking at $t=8.5 \pm 1.1$ days after GRB 100316D \citep{Cano2011,Bufano2012ApJ,Olivares2012}, which is significantly later than that found here.
Therefore, in addition to the remarkably soft spectrum at about 13.6 days after the burst noticed by Y15,
the rather early peak of the excess components found in this work strongly disfavors an SN interpretation. In the NS-NS merger scenario, \citet{Metzger2014} found that some regions of the outflow may be Lanthanide-free, and such material will become optically thin within a few days after the merger and give rise to optical/UV emission lasting for a day or so.  Such a scenario can, at least in part, explain the early peak in the macronova observed here.  A successful interpretation of the very significant F814W-band excess emission at $t\sim 13.6$ day is, however, a challenge for theoretical NS-NS merger models.

\begin{figure}
\begin{center}
 \includegraphics[width=0.5\textwidth]{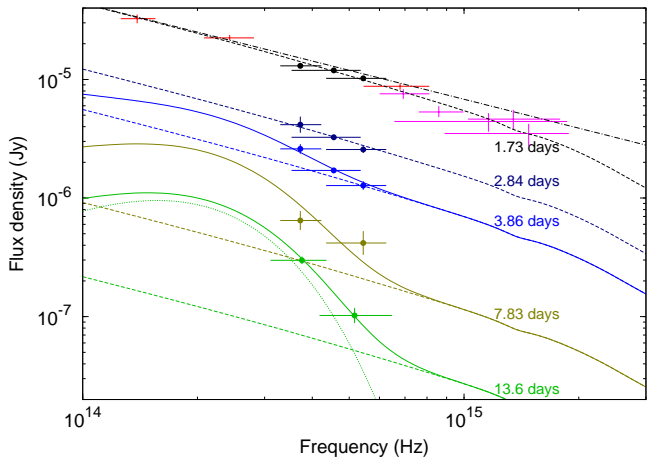}
\end{center}
\caption{The observed SED evolution of GRB 060614.
From top to bottom are the SEDs at $t=(1.73,~2.84,~3.86,~7.83,~13.6)$ days, respectively.
Solid circles are from Y15, red crosses are VLT data from \citet{DellaValle2006} and purple crosses are Swift UVOT data from \citet{Mangano2007},
all have not been corrected for extinction.
In two early observations the SEDs can be fitted with a single power-law spectrum with extinction of the Galaxy and the host galaxy,
where dash dot line is the intrinsic spectrum and dashed lines are extincted.
The remaining three observations are fitted by a single power-law and a blackbody spectrum ($T=2700$ K, dotted line),
where extinction has been taken into account.
}
\label{fig:SED}
\end{figure}

There are five epochs that consist of two or more filters,
which we have combined into five spectral energy distributions (SEDs), and are displayed in  Fig. \ref{fig:SED}.
An excess of flux (i.e., relative to a single power-law spectrum) is clearly visible in the three latter epochs.
Instead, blackbody spectra provide a reasonable fit to the observed SEDs (see Fig. \ref{fig:SED}).
At $t\sim 13.6$ days after the burst, the temperature is estimated to be $2700_{+700}^{-500}$ K.
At other times, the temperatures are poorly constrained (i.e., $<4200$ K and $3100-19000$ K at 3.86 and 7.83 days respectively).
Due to their large errors, it is impossible to draw any conclusions regarding a possible temperature evolution.

As pointed out in Y15, the progenitor system was likely a BH-NS binary, as these types of merger are expected to give rise to ``bluer", longer and brighter macronova emission than NS-NS mergers due to more ejecta mass and a highly-anisotropic distribution of the ejecta material \citep[see][and the references therein]{Tanaka2014,Kyutoku2015}.
For GRB 060614, to account for the distinct F814W-band excess at $t\sim 13.6$ days,
a simple estimate based on the generation of the macronova lightcurve in one BH-NS merger model presented in \citet{Tanaka2014}
suggests that the ejected material from the merger was $\sim 0.1~M_\odot$ and moved at a velocity $\sim 0.2c$.
In this paper, the F814W-band excess is just a bit brighter and the parameters of the ejecta are likely similar to those in previous estimate.
The peak emission of VLT/$I$-band ($R$-band) excess is as bright as $\sim 24^{\rm th}$ mag ($\sim 25^{\rm th}$ mag),
in agreement  with the merger macronova model (see Fig. \ref{fig:LC}).
A reliable interpretation of the macronova lightcurve, however,
requires dedicated numerical simulation studies including a proper convolution of the produced complicated macronova spectra with the response function of the most widely used facilities in order to aid the comparison with actual data,
which is beyond the scope of this work.

\begin{table}
\label{tab:macronova}
\begin{center}
\title{}Table 1. The macronova component of GRB 060614\\
\begin{tabular}{lllll} \hline \hline
Time from GRB & Filter & Magnitude$^{a}$\\
(days)    &	& (Vega)  \\ \hline \hline
7.828   & VLT $V$ & (25.6$\pm$0.6)\\
3.869   & VLT $R$ & (25.3$\pm$0.6)\\
4.844   & VLT $R$  & 24.9$\pm$0.3\\
6.741   & VLT $R$ & 25.3$\pm$0.3\\
10.814  & VLT $R$ & (26.5$\pm$0.8)\\
14.773   & VLT $R$ & (27.2$\pm$1.0)\\
3.858 & VLT $I$ & 23.7$\pm$0.4\\
7.841 & VLT $I$ & 24.6$\pm$0.4\\
13.970 & HST F606W & 26.9$\pm$0.4 \\
13.571 & HST F814W & 25.05$\pm$0.12 \\
\hline
\end{tabular}
\end{center}
Note: a. The magnitudes of the extracted macronova component.
The observations with errors larger than 0.5 mag have been bracketed.
\end{table}

\section{The rate of the macronova and compact-object mergers}
So far, two macronovae have been observed at redshifts of $z=0.356$ and  $z=0.125$ for GRB 130603B \citep{Tanvir2013,Berger2013} and GRB 060614 (Y15), respectively. Both events were found to be collimated with a half-opening angle $\theta_{\rm j}\sim 0.1$ \citep[e.g.,][]{Xu2009,Fan2013,Fong2014}. For the macronovae at $z\geq 0.4$, HST observations are crucial to get the signal but HST observations of such ``high"-$z$ short GRBs  were very rare \citep{Berger2014}. The current sample can be taken as that recordable by {\it Swift}, an instrument with a field-of-view of 2 steradians, in the last ten years for the events with $z\leq 0.4$.
With these numbers in mind, we estimated the local macronova/compact-object merger rate to be  \[{\cal R}_{\rm macronova}\sim 16.3^{+16.3}_{-8.2}~{\rm Gpc^{-3}~yr^{-1}}(\theta_{\rm j}/0.1)^{-2}.\] Note that this rate is corrected for beaming, as such it is compatible with the un-beamed SGRB rate of $4\pm 2 ~{\rm Gpc^{-3}~yr^{-1}}$ \citep{Wanderman2014}. For the  upcoming Advanced LIGO/VIRGO/KAGRA detectors that can detect the gravitational wave radiation from compact-object mergers within a distance  $D\sim 200$ Mpc \citep{Aasi2013}, the detection rate is expected to be \[{\cal R}_{\rm GW} \sim 0.5^{+0.5}_{-0.25}(D/200~{\rm Mpc})^{3}~{\rm yr^{-1}}.\] Bear in mind that such rates are (very) conservative since (1) macronova searches usually need HST-like detectors and not all SGRBs and long-short GRBs have been followed down to deep limits; (2) it is likely that just a fraction of compact-object mergers can produce GRBs. Hence the above estimates are better taken as lower limits. We thus conclude that the prospect of detecting gravitational wave radiation from merger events in the near future is quite promising.

Interestingly, a realistic estimate of the BH-NS merger rate is $\sim 30~{\rm Gpc^{-3}}{\rm yr^{-1}}$ \citep{Abadie2010}, which is compatible with ${\cal R}_{\rm macronova}$ estimated here, implying that a BH-NS merger origin for GRB 060614 is indeed plausible.

\section{Discussion}
Since \citet{Li1998} first proposed that there may be a near-infrared/optical transient following the merger of a compact binary,
significant progress has been made in numerical simulations \citep[e.g.,][and the references therein]{Barnes2013,Kasen2013,Tanaka2013,Tanaka2014,Kyutoku2015}.  Conversely, observational macronova signatures have only been detected for SGRB 130603B \citep{Tanvir2013,Berger2013} and long-short burst GRB 060614 \citep{Yang2015}.

Due to the lack of (detailed) lightcurves and spectra, the knowledge we can learn is rather limited and the predictions made in the numerical simulations can not be fully tested. For example, \citet{Hotokezaka2013} showed that for the single macronova data point of SGRB 130603B the NS-NS and BH-NS merger scenarios can not be distinguished. In the present work, with the assumption that the afterglow data in the time interval of $1.7-3.0$~days after GRB 060614 are generated by external forward shock we have shown that at late times there are significant excess components in multi-wavelength photometric observations (see Fig. \ref{fig:LC}).
There is evidence shows that the associated macronova likely peaked at $t\lesssim 4$ days after the $\gamma-$ray transient, which is consistent with current numerical simulations of macronova emission but much earlier than the peak times of GRB-associated SNe.

In the approximation of a thermal spectrum, the temperature of the excess component is inferred to be $\sim 2700$ K at $t\sim 13.6$ day. Due to the limited data, no strong evidence for evolution of the temperature can be established. Such a temperature is significantly lower than that of an SN at the same time scale, typically $\sim 0.5-1\times 10^{4}$ K (see e.g., \citealt{DellaValle2006} and \citealt{Cano2011}), but it is similar to that expected at the photosphere for the recombination of Lanthanides \citep[i.e., $T\sim2500$ K, see e.g., ][]{Barnes2013}. This lends additional support to the neutron-rich nature of macronova ejecta.

We conservatively estimated the macronova rate ${\cal R}_{\rm macronova}\sim 16.3^{+16.3}_{-8.2}~{\rm Gpc^{-3}}{\rm yr^{-1}}$, and implied that the detection prospect of the gravitational wave radiation from compact object mergers by upcoming Advanced LIGO/VIRGO/KAGRA detectors is quite promising, where the expected rate is ${\cal R}_{\rm GW} \sim 0.5^{+0.5}_{-0.25}(D/200~{\rm Mpc})^{3}~{\rm yr^{-1}}$.

In the foreseeable future, it is anticipated that increasingly more macronova lightcurves will be recorded. There could be two types of macronova lightcurves at least. One group is from NS-NS mergers and the other is from BH-NS mergers. On one hand, BH-NS mergers are expected to give rise to ``bluer", longer and brighter macronova emission than the NS-NS mergers \citep[e.g.,][]{Hotokezaka2013}. On the other hand, the BH-NS merger rate is generally expected to be at most $\sim 1/10$ time that of the NS-NS merger rate \citep{Abadie2010}. Hence, if the macronova associated with GRB 060614 indeed arose from a BH-NS merger, its lightcurve will be different from the majority of the sample that is expected to be from NS-NS mergers.

\section*{Acknowledgments}
We thank the referee for insightful comments. This work was supported in part by National Basic Research Programme of China (No. 2013CB837000 and No. 2014CB845800),
NSFC under grants 11361140349, 11103084, 11273063, 11433009 and U1231101,
the Foundation for Distinguished Young Scholars of Jiangsu Province, China (Grant No. BK2012047)
and the Strategic Priority Research Program (Grant No. XDB09000000).
Z.C. is funded by a Project Grant from the Icelandic Research Fund.
S.C. has been supported by ASI grant I/004/11/0.\\

\clearpage


\begin{thebibliography}{}

\bibitem[Aasi et al. (2013)]{Aasi2013} Aasi, J., Abadie, J., Abbott, B. P., et al. (LIGO Scientific Collaboration, Virgo Collaboration) 2013, arXiv:1304.0670

\bibitem[Abadie et al. (2010)]{Abadie2010} Abadie, J., Abadie, J., Abbott, B. P., et al. 2010, CQGra, 27, 173001

\bibitem[Barnes \& Kasen (2013)]{Barnes2013} Barnes, J. \& Kasen, D. 2013, ApJ, 773, 18.

\bibitem[Berger et al. (2013)]{Berger2013} Berger, E., Fong, W., \& Chornock, R. 2013, ApJL, 744, L23

\bibitem[Berger (2014)]{Berger2014} Berger, E., 2014, ARA\&A, 52, 43

\bibitem[Bufano et al. (2012)]{Bufano2012ApJ} Bufano, F., Pian, E., Sollerman, J., et al. 2012, ApJ, 753, 67

\bibitem[Cano et al. (2011)]{Cano2011} Cano, Z., Bersier, D., Guidorzi, C., et al. 2011, ApJ, 740, 41

\bibitem[Della Valle et al. (2006)]{DellaValle2006} Della Valle, M., Chincarini, G., Panagia, N.,  et al. 2006, Natur, 444, 1050

\bibitem[Eichler et al. (1989)]{Eichler1989} Eichler D.,  Livio M.,  Piran T., \& Schramm D. N. 1989, Natur, 340, 126

\bibitem[Evans et al. (2009)]{Evans2009} Evans, P. A., Beardmore, A. P., Page, K. L., et al. 2009, MNRAS, 397, 1177

\bibitem[Fan et al. (2013)]{Fan2013} Fan, Y. Z., Yu, Y. W., Xu, D., et al. 2013, ApJL, 779, L25

\bibitem[Fong et al. (2010)]{Fong2010} Fong, W., Berger, E., \& Fox, D. B. 2010, ApJ, 708, 9

\bibitem[Fong \& Berger (2013)]{Fong2013} Fong, W., \& Berger, E., 2013, ApJ, 776, 18

\bibitem[Fong et al. (2014)]{Fong2014} Fong, W., Berger, E., Metzger, B. D., et al. 2014, ApJ, 780, 118

\bibitem[Fynbo et al. (2006)]{Fynbo2006} Fynbo, J. P. U., Watson, D., Th\"one, C. C., et al. 2006, Natur, 444, 1047

\bibitem[Gal-Yam et al. (2006)]{Gal-Yam2006} Gal-Yam, A., Fox, D. B., Price, P. A., et al. 2006, Natur, 444, 1053

\bibitem[Gehrels et al. (2006)]{Gehrels2006} Gehrels, N., Norris, J. P., Barthelmy, S. D., et al. 2006, Natur, 444, 1044

\bibitem[Gehrels et al. (2005)]{Gehrels2005} Gehrels, N., Sarazin, C. L., O'Brien, P. T., et al. 2005, Natur, 437, 851

\bibitem[Grossman et al. (2014)]{Grossman2014} Grossman, D., Korobkin, O., Rosswog, S., \& Piran, T. 2014, MNRAS, 439, 757

\bibitem[Hotokezaka et al. (2013)]{Hotokezaka2013} Hotokezaka, K.,  Kyutoku, K., Tanaka, M., et al. 2013, ApJL, 778, L16

\bibitem[Kasen et al. (2013)]{Kasen2013} Kasen, D., Badnell, N. R. \& Barnes, J. 2013, ApJ, 774, 25

\bibitem[Kisaka et al. (2015a)]{Kisaka2015a} Kisaka, S., Ioka, K., Takami, H. 2015a, ApJ, 802, 119

\bibitem[Kisaka et al. (2015b)]{Kisaka2015b} Kisaka, S., Ioka, K., Takami, H. 2015b, arXiv:1506.02030

\bibitem[Korobkin et al. (2012)]{Korobkin12} Korobkin, O., Rosswog, S.,  Arcones, A., \& Winteler, C. 2012, MNRAS, 426, 1940

\bibitem[Kulkarni (2005)]{Kulkarni2005} Kulkarni, S. R. 2005, arXiv:astro-ph/0510256

\bibitem[Kumar \& Zhang (2015)]{Kumar2015} Kumar, P., \& Zhang, B., 2015, PhR, 561, 1

\bibitem[Kyutoku et al. (2015)]{Kyutoku2015} Kyutoku, K., Ioka, K., Okawa, H., Shibata, M., \& Taniguchi, K. 2015, arXiv:1502.05402

\bibitem[Leibler et al. (2010)]{Leibler2010} Leibler C. N., \& Berger E. 2010, ApJ, 725, 1202

\bibitem[Li \& Paczynski (1998)]{Li1998} Li, L.-X., \& Paczy\'{n}ski, B. 1998, ApJL, 507, L59

\bibitem[Lippuner \& Roberts (2015)]{Lippuner2015} Lippuner, J. \& Roberts, L. F. 2015, arXiv:1508.03133

\bibitem[Mangano et al. (2007)]{Mangano2007} Mangano, V.,  Holland, S. T., Malesani, D. et al. 2007, A\&A, 470, 105

\bibitem[Metzger et al. (2010)]{Metzger2010} Metzger, B. D., Mart\'{i}nez-Pinedo, G., Darbha, S. et al. 2010, MNRAS, 406, 2650

\bibitem[Metzger \& Fern\'{a}ndez (2014)]{Metzger2014} Metzger, B. D. \& Fern\'{a}ndez, R. 2014, MNRAS, 441, 3444

\bibitem[Narayan et al. (1992)]{Narayan1992} Narayan, R., Paczynski, B., \& Piran, T. 1992,  ApJL, 395, L83

\bibitem[Olivares et al. (2012)]{Olivares2012} Olivares, E. F., Greiner, J., Schady, P. et al. 2012, A\&A, 539, A76

\bibitem[Piran (2004)]{Piran2004} Piran, T., 2004, RvMP, 76, 1143

\bibitem[Piran et al. (2014)]{Piran2014} Piran, T., Korobkin, O., Rosswog, S., 2014, arXiv:1401.2166

\bibitem[Rosswog (2005)]{Rosswog2005} Rosswog, S. 2005, ApJ, 634, 1202

\bibitem[Sari et al. (1999)]{Sari1999} Sari, R. Piran, T. \& Halpern, 1999, ApJL, 519, L17

\bibitem[Schlafly \& Finkbeiner(2011)]{2011ApJ...737..103S} Schlafly, E.~F., \& Finkbeiner, D.~P.\ 2011, ApJ, 737, 103

\bibitem[Schlegel, Finkbeiner \& Davis(1998)]{Schlegel98} Schlegel, D. J., Finkbeiner, D. P., \& Davis, M.\ 1998, ApJ, 500, 525

\bibitem[Tanaka \& Hotokezaka (2013)]{Tanaka2013} Tanaka, M., \& Hotokezaka, K. 2013, ApJ, 775, 113

\bibitem[Tanaka et al. (2014)]{Tanaka2014} Tanaka, M., Hotokezaka, K., Kyutoku, K. et al. 2014, ApJ, 780, 31

\bibitem[Tanvir et al. (2013)]{Tanvir2013} Tanvir, N. R., Levan, A. J., Fruchter, A. S. et al. 2013, Natur, 500, 547

\bibitem[Wanajo et al. (2014)]{Wanajo2014} Wanajo, S., Sekiguchi, Y., Nishimura, N. et al. 2014, ApJL, 789, L39

\bibitem[Wanderman \& Piran (2015)]{Wanderman2014} Wanderman, D., \& Piran, T. 2015, MNRAS, 448, 3026

\bibitem[Xu et al. (2009)]{Xu2009} Xu, D., Starling, R. L. C., Fynbo, J. P. U. et al. 2009, ApJ, 696, 971

\bibitem[Yang et al. (2015)]{Yang2015} Yang, B., Jin, Z. P., Li, X. et al. 2015, Nat. Commun., 6, 7323 (Y15)

\bibitem[Zhang et al.(2007)]{Zhang2007} Zhang, B., Zhang, B. B., Liang, E. W. et al. 2007, ApJL, 655, L25

\end{thebibliography}
\end{document}